\documentclass[prd,aps,nofootinbib,10pt]{revtex4}
\usepackage{amsmath,graphicx,epsfig,amssymb,dsfont,mathtools}

\usepackage[usenames]{color}
\usepackage{ulem} 
\usepackage{bigstrut}

\begin{document}
\title{Chiral Dynamics and S-wave contributions in Semileptonic $D_s/B_s$   decays into $\pi^+\pi^-$ }
\author{Yu-Ji Shi$^1$~\footnote{Email:{shiyuji@sjtu.edu.cn}}, and  Wei Wang$^{1,2}$~\footnote{Email:{wei.wang@sjtu.edu.cn}}}

\affiliation{
$^1$ INPAC, Shanghai Key Laboratory for Particle Physics and Cosmology, Department of Physics and Astronomy, Shanghai Jiao-Tong University, Shanghai, 200240,   China\\
$^2$
State Key Laboratory of Theoretical Physics, Institute of Theoretical Physics, Chinese Academy of Sciences, Beijing 100190, China}

\begin{abstract}
In this work, we study the semileptonic decay modes $B_s^0\to \pi^+\pi^-\ell^+\ell^-$ and  $D_s^+\to \pi^+\pi^-\ell^+ \nu$   in the kinematics  region where the $\pi^+\pi^-$ system  has a invariant mass in the range $0.5$-$1.3$ GeV.  These  processes  are   valuable towards the determination of   S-wave $\pi^+\pi^-$ light-cone distribution amplitudes whose normalizations are scalar  form factors. We compare the results for scalar form factors predicted in  unitarized chiral perturbation theory and extracted from the data on the $B_s\to J/\psi \pi^+\pi^-$.  Then the $B_s\to \pi^+\pi^-$ and $D_s\to \pi^+\pi^-$ form factors are calculated in light-cone sum rules, based on which predictions   for differential decay widths are made. The results  are in good agreement   with the experimental data on the $B_s$ and $D_s$ decays into $\pi^+\pi^-$.  More accurate measurements at BEPC, LHC and KEKB  in future will be helpful  to examine our formalism and constrain the input parameters more precisely.
\end{abstract}
\maketitle


\section{Introduction}


It is anticipated that    new physics (NP) beyond the standard model (SM) can be  indirectly probed through the precision exploration of low-energy processes. An  ideal platform is to study the flavor-changing neutral current (FCNC). Rare $B$ decays like  the $b\to s\ell^+\ell^-$, with tiny
decay probabilities in the SM, are  sensitive to   NP
degrees of freedom and thus can be exploited as indirect searches
of these unknown effects.  In terms of  observables ranging
from the decay probabilities, forward-backward asymmetries,
polarizations to a full angular analysis, the exclusive decay mode $B\to K^*\ell^+\ell^-$ can provide us with a wealth of information on weak interactions.   Recent measurements of the almost form-factor independent ratio $P_5'$ by the LHCb collaboration  have indicated a deviation from the SM by about $3.7\sigma$~\cite{Aaij:2013qta,LHCb:2015dla}.

In fact, the $B\to K^*\ell^+\ell^-$ is a four-body decay process since the $K^*$ meson is reconstructed in the $K\pi$ final state. Thus it is more appropriate to explore the $B\to M_1M_2\ell^+\ell^-$, in which various partial-waves of $M_1M_2$ contribute~\cite{Lu:2011jm,Dey:2015rqa,Gratrex:2015hna}. The S-wave contributions to $B\to K\pi\ell^+\ell^-$ have been discussed for instance  in Refs.~\cite{Becirevic:2012dp,Matias:2012qz,Blake:2012mb,Meissner:2013hya,Das:2014sra,Hofer:2015kka,Das:2015pna}.  The bottom  mass $m_b$ is much heavier than the hadronic scale $\Lambda_{\rm QCD}$, which allows an expansion of  the hard-scattering kernels in terms of the strong coupling constant $\alpha_s$ and the power-scaling  parameter $\Lambda_{\rm QCD}/m_b$.  On the other side, final state interactions among the two-light hadrons should be  constrained by unitarity and analyticity.
A formalism that makes use of  these two advantages has been developed in Refs.~\cite{Meissner:2013hya,Meissner:2013pba,Doring:2013wka}, and  summarized in Ref.~\cite{Wang:2014sba}. Such an approach  was pioneered in Ref.~\cite{Gardner:2001gc,Maul:2001zn},  and a method without  the analysis  of    hard-scattering kernels has been explored  recently  in Refs.~\cite{Liang:2014tia,Bayar:2014qha,Xie:2014gla,Sayahi:2013tza,Roca:2015tea,Sekihara:2015iha}. See also Refs.~\cite{Chen:2002th,Chen:2004az,Wang:2014ira,Wang:2014qya} for attempts to analyze charmless three-body $B$ decays.
The aim of this work is to further examine this formalism by confronting the theoretical results  with  the relevant  data on $B_s\to\pi^+\pi^-\mu^+\mu^-$  and $D_s\to  \pi^+\pi^- e^+ \nu_{e}$.

In Ref.~\cite{Aaij:2014lba}, the LHCb collaboration  has performed an analysis of rare $B_s$ decays into the $\pi^+\pi^-\mu^+\mu^-$ final state with the measured   branching ratio:
\begin{eqnarray}
{\cal B}(B_s\to  \pi^+\pi^- \mu^+\mu^-) = (8.6\pm 1.5\pm 0.7\pm 0.7)\times 10^{-8}, \label{eq:Bspipidata}
\end{eqnarray}
where the errors are statistical,   systematic and  arise from the normalization, respectively.   The dominant contribution is    the $B_s\to f_0(980) \mu^+\mu^-$~\cite{Aaij:2014lba}:
\begin{eqnarray}
{\cal B}(B_s\to f_0(980)(\to \pi^+\pi^-) \mu^+\mu^-) = (8.3\pm 1.7)\times 10^{-8}. \label{eq:Bs_f0_mumu_data}
\end{eqnarray}
This has triggered theoretical interpretations  based on two-meson light-cone distribution amplitudes (LCDAs)~\cite{Wang:2015uea,Wang:2015paa}.

Previously, the CLEO collaboration has investigated the $D_s\to \pi^+\pi^-\ell \nu_{e}$, in which the $f_0(980)$ contribution  is found dominant as well~\cite{Yelton:2009aa,Ecklund:2009aa}
\begin{eqnarray}
{\cal B}(D_s\to f_0(980)(\to \pi^+\pi^-) e^+ \nu_{e})= (2.0\pm0.3\pm0.1)\times 10^{-3}. \label{eq:Ds_f0_pipi_CLEO_c}
\end{eqnarray}
A recent analysis based on the CLEO-c data~\cite{Hietala:2015jqa} gives a similar result
\begin{eqnarray}
{\cal B}(D_s\to f_0(980)(\to \pi^+\pi^-) e^+ \nu_{e})= (1.3\pm0.2\pm0.1)\times 10^{-3}. \label{eq:Ds_f0_pipi_CLEO_c_2}
\end{eqnarray}
The BES-III collaboration  will collect about $2fb^{-1}$ data in $e^+e^-$ collision  at the energy around $4.17$GeV,  which will be  used to study semileptonic and nonleptonic $D_s$ decays~\cite{Asner:2008nq}.


The rest of this paper is organized as follows.  In Sec.~\ref{sec:pipi_Scalar}, we will give the results for  scalar $\pi\pi$ form factors and the non-local LCDAs.  Sec.~\ref{sec:heavy_to_light} will  be devoted to the calculation of the $B_s\to \pi^+\pi^-$ and $D_s\to \pi^+\pi^-$ form factors in the light-cone sum rules (LCSR). In Sec.~\ref{sec:phono},   phenomenological results for a variety of  observables in the $B_s\to \pi^+\pi^-\ell^+\ell^-$, $B_s\to \pi^+\pi^-\nu\bar\nu$ and $D_s\to \pi^+\pi^-\ell\nu$ are presented, and compared to the experimental data if available.  An agreement between   theory and    data will be shown in this section. Our conclusions will be  given in Sec.~\ref{sec:conclusion}.

\section{Scalar $\pi^+\pi^-$ Form Factors and S-wave  LCDAs}
\label{sec:pipi_Scalar}

\subsection{Scalar form factor}

We start with the definition of a scalar  form factor:
\begin{eqnarray}
\langle \pi^+\pi^-| \bar ss|0
\rangle= B_0\, F_{\pi\pi}^{s}(m_{\pi\pi}^2),
\end{eqnarray}
and the $B_0$ is the QCD condensate parameter:
\begin{eqnarray}
 \langle 0|\bar qq|0\rangle \equiv- f_{\pi}^2 B_0,
\end{eqnarray}
with $f_{\pi}$   as the leading order (LO) pion decay constant. For the numerics, we use $f_{\pi}=91.4$MeV and
$ \langle 0|\bar qq|0\rangle =-[(0.24\pm0.01){\rm GeV}]^3$~\cite{Colangelo:2000dp}. This corresponds to $B_0=(1.7\pm0.2)$ GeV.

\begin{figure}\begin{center}
\includegraphics[scale=0.6]{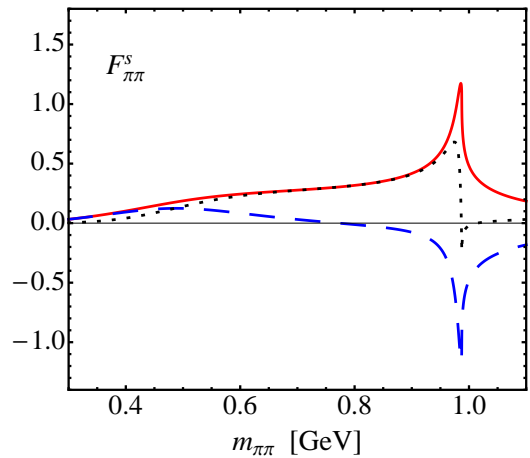}
\caption{The $F_{\pi\pi}^{s}(m_{\pi\pi}^2)$     in   unitarized $\chi$PT. The modulus, real part and imaginary part are shown in solid, dashed and dotted curves. The result  is based on the fit of the BES data on the $\pi\pi$ invariant mass distributions in $J/\psi\to\pi^+\pi^-\phi$~\cite{Ablikim:2004wn}  in Ref.~\cite{Lahde:2006wr}.  } \label{fig:pipi_ff}
\end{center}
\end{figure}

In the literature,   a variety of approaches
have been used to calculate the $F^s_{\pi\pi}(m_{\pi\pi}^2)$, including   the (unitarized) chiral perturbation
theory ($\chi$PT)~\cite{Gasser:1990bv,Meissner:2000bc,Bijnens:2003uy,Lahde:2006wr,Guo:2012yt,Gasser:1983yg,Gasser:1984gg,Gasser:1984ux}
and dispersion
relations~\cite{Donoghue:1990xh}.  In the $\chi$PT,  the LO and next-to-leading order (NLO) results can describe the low-energy data  with a good accuracy, but with the increase of the invariant mass, higher-order contributions become more important.   This is under expectation since the perturbative  expansion in  $\chi$PT is organized in terms of $p_{\pi}/(4\pi f_\pi)$.
It has been argued  that the unitarized approach, by summing  higher order corrections, can extend the $\chi$PT applicability to the scale  around 1 GeV~\cite{Oller:1998hw}.
A fit to the BES data on the $\pi\pi$ invariant mass distributions in $J/\psi\to\pi^+\pi^-\phi$~\cite{Ablikim:2004wn} has been performed  in  this approach and an overall agreement is found~\cite{Lahde:2006wr}.  The fitted result for $F^s_{\pi\pi}(m_{\pi\pi}^2)$ is shown  in Fig.~\ref{fig:pipi_ff}.  The modulus, real part and
imaginary part are shown as solid, dashed and dotted curves, from which one can observe  the broad structure for the $\sigma(600)$ and the peak at $f_0(980)$ is naturally produced.

In recent years the LHCb collaboration has conducted  a series of  analyses of angular distributions in the $B_s\to J/\psi\pi^+\pi^-$ decay mode~\cite{Aaij:2013zpt,Aaij:2014emv} .  In this procedure,  the S-wave contributions have been explicitly  separated, and  three resonances, $f_0(980)$, $f_0(1500)$ and $f_0(1790)$, have been identified.
To access the $\pi^+\pi^-$ invariant mass distribution, the Breit-Wigner formula is employed for the $f_0(1500)$ and $f_0(1790)$.
Due to the fact that  the $f_0(980)$ lies in the vicinity of the $K\bar K$ threshold,
the Flatt\'e model~\cite{Flatte:1976xu,Flatte:1976xv} has been adopted.    Considering the relative strengths and strong phases among different resonances,
we have
\begin{eqnarray}
F_{\pi\pi}^{s}(m_{\pi\pi}^2)&=&
\frac{c_1m_{f_0(980)}^2e^{i\theta_1}}{m_{\pi\pi}^2-m_{f_0(980)}^2+im_{f_0(980)}
  (g_{\pi\pi}\rho_{\pi\pi}+g_{KK}F_{KK}^2\rho_{KK})} \nonumber\\
  &&+\frac{c_2m_{f_0(1500)}^2e^{i\theta_2}}{m_{\pi\pi}^2-m_{f_0(1500)}^2+im_{f_0(1500)}\Gamma_{f_0(1500)}(m_{\pi\pi}^2)}\nonumber\\
& &+\frac{c_3m_{f_0(1790)}^2e^{i\theta_3}}{m_{\pi\pi}^2-m_{f_0(1790)}^2+im_{f_0(1790)}\Gamma_{f_0(1790)}(m_{\pi\pi}^2)}.
\label{eq:ff-exp_data}
\end{eqnarray}
The  $\rho_{\pi\pi}$ and $\rho_{KK}$ are  phase space factors~\cite{Aaij:2013zpt,Flatte:1976xv,Aaij:2014emv}:
\begin{eqnarray}
\rho_{\pi\pi}=\frac23\sqrt{1-\frac{4m^2_{\pi^\pm}}{m_{\pi\pi}^2}}
 +\frac13\sqrt{1-\frac{4m^2_{\pi^0}}{m_{\pi\pi}^2}},\quad
\rho_{KK}=\frac12\sqrt{1-\frac{4m^2_{K^\pm}}{m_{\pi\pi}^2}}
 +\frac12\sqrt{1-\frac{4m^2_{K^0}}{m_{\pi\pi}^2}}.
\end{eqnarray}
Compared to the normal Flatt\'e distribution, an additional correction  has been introduced in  the LHCb fit above the $K\bar K$ threshold to better describe the data~\cite{Aaij:2014emv}
\begin{eqnarray}
 F_{KK}= {\rm exp} (-\alpha k^2),
\end{eqnarray}
where $k$ is the kaon momentum   in the $K\bar K$ rest frame, and $\alpha$ is set to $\alpha=2.0 {\rm GeV}^{-2}$~\cite{Aaij:2014emv}.
The energy-dependent
width $\Gamma_S(m_{\pi\pi}^2)$ for an $S$-wave resonance  is parameterized as
\begin{eqnarray}
\Gamma_S(m_{\pi\pi}^2)=\Gamma_S \frac{m_S}{m_{\pi\pi}}
\left(\frac{m_{\pi\pi}^2-4m_{\pi}^2}{m_S^2-4m_{\pi}^2} \right)^{\frac{1}{2}} F_R^2,
\end{eqnarray}
with  the constant width $\Gamma_S$, and the Blatt-Weisskopf
barrier factor $F_R=1$~\cite{Aaij:2014emv}.  The
$c_i$ and $\theta_i$, $i=1$, 2, and 3, are tunable parameters.   In the fit by LHCb, two solutions are found and the fitted parameters for three contributing components are collected in Tab.~\ref{tab:fitted_para}~\cite{Aaij:2014emv}.

\begin{table}[t]
\begin{center}
\caption{Fitted parameters for  contributing components in the $B_s\to J/\psi \pi^+\pi^-$ by the LHCb collaboration~\cite{Aaij:2014emv}. Two solutions are found in the fit.  }
\begin{tabular}{lcc}
\hline
 Fractions (\%)            & Solution I & Solution II  \\\hline
$f_0(980)$ & $70.3\pm1.5_{-5.1}^{+0.4}$ & $92.4\pm2.0_{-16.0}^{+~0.8}$\\
$f_0(1500)$ & $10.1\pm0.8_{-0.3}^{+1.1}$ & $9.1\pm0.9\pm0.3$\\
$f_0(1790)$ & $2.4\pm0.4_{-0.2}^{+5.0}$ & $0.9\pm0.3_{-0.1}^{+2.5}$\\
\hline
 Phase differences ($^\circ$)           & Solution I & Solution II  \\\hline
$f_0(1500)-f_0(980)$ &$138\pm4$ & $177\pm6$\\
$f_0(1790)-f_0(980)$ & $78\pm9$& $95\pm16$\\ \hline
Parameter & Solution I& Solution II\\\hline
$m_{f_0(980)}$ ({\rm MeV})&$945.4\pm2.2$&$949.9\pm2.1$\\
$g_{\pi\pi}$~({\rm MeV})&$167\pm7$&$167\pm8$\\
$g_{KK}/g_{\pi\pi}$&$3.47\pm0.12$&$3.05\pm0.13$\\
$m_{f_0(1500)}$ ({\rm MeV})&$1460.9\pm2.9$&$1465.9\pm3.1$\\
$\Gamma_{f_0(1500)}$ ({\rm MeV})&$124\pm7$&$115\pm7$\\
$m_{f_0(1790)}$ ({\rm MeV})&$1814\pm18$&$1809\pm22$\\
$\Gamma_{f_0(1790)}$ ({\rm MeV})&$328\pm34$&$263\pm30$\\\hline
\hline
\end{tabular}
\label{tab:fitted_para}
\end{center}
\end{table}

\begin{figure}\begin{center}
\includegraphics[scale=0.5]{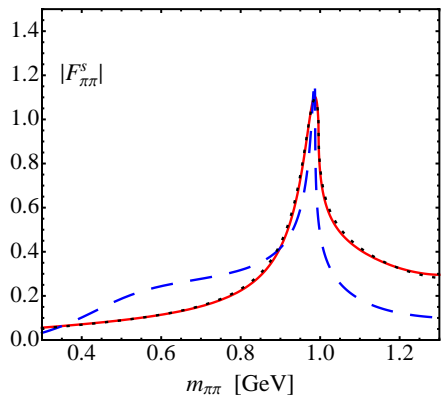}
\caption{The  $F_{\pi\pi}^{s}(m_{\pi\pi}^2)$ predicted  in the $\chi$PT (dashed curve), and fitted  from the $B_s\to J/\psi\pi^+\pi^-$ data (solid line for solution and dotted line for solution II).  The two fitted  solutions are not distinguishable below $1.2$ GeV.  } \label{fig:pipi_ff_compa}
\end{center}
\end{figure}

In Fig.~\ref{fig:pipi_ff_compa},  we compare  the results for the scalar form factor calculated  in the unitarized $\chi$PT (dashed curve), and the ones extracted from the $B_s\to J/\psi\pi^+\pi^-$ data based on  Eq.~\eqref{eq:ff-exp_data} (solid and dotted curves). The constant $c_1$  in  Eq.~\eqref{eq:ff-exp_data} has been tuned in the comparison.
The two solutions found by the LHCb give very similar shapes as shown in the figure.
There are a few remarks on the shapes.

\begin{itemize}
\item On the contrary with the $\chi$PT results, the parametrization   in Eq.~(\ref{eq:ff-exp_data}) does not contain the $f_0(500)$ (or the so-called $\sigma$) contribution. The LHCb collaboration has set an upper limit for the ratio~\cite{Aaij:2014emv}:
\begin{eqnarray}
R= \frac{F_r(f_0(500))}{F_r(f_0(980)}< 0.3\%
\end{eqnarray}
at a $90\%$ C.L.

\item The expansion parameter in   $\chi$PT is $p_{\pi}/(4\pi f_{\pi})$, where the $p_\pi$ is the pion's momentum. Summing higher  order s-channel contributions and incorporating the coupled channel effects will extend the applicability region up to 1GeV.  On the other hand, the parametrization in Eq.~(\ref{eq:ff-exp_data})  has explicitly  included    two scalar resonances, $f_0(1500)$ and $f_0(1790)$, and thus  is valid in the region above 1 GeV.

\item At the $f_0(980)$, the unitarized $\chi$PT leads to a  very narrow peak. This feature is smeared out in the data since the binned results for the $\pi\pi$ invariant mass distributions  are presented on the experimental side. Still  the parametrization  in Eq.~(\ref{eq:ff-exp_data}) is well consistent with the data on  the $B_s\to J/\psi\pi^+\pi^-$~\cite{Aaij:2013zpt,Aaij:2014emv}.

\item  From the comparison, we can see that at the current stage the advantages in both cases  are not overwhelming. In future, we  hope the results can be improved in  some more sophisticated methods like  the unitarized approach with resonances~\cite{Guo:2012yt}.  In the following, we will use the parametrization  form  inspired by the data in  Eq.~\eqref{eq:ff-exp_data}.

\end{itemize}

\subsection{Generalized LCDAs}

With the scalar form factor as the normalization condition,
the S-wave $\pi^+\pi^-$  LCDAs   are defined by~\cite{Diehl:1998dk,Polyakov:1998ze,Kivel:1999sd,Diehl:2003ny,Meissner:2013hya}:
\begin{eqnarray}
 \langle {(\pi^+\pi^-)_S}|\bar s (x)\gamma_\mu s(0)|0\rangle &=&  F_{\pi\pi}^{s}(m_{\pi\pi}^2)
 p_{{\pi\pi},\mu}
 \int_0^1 due^{i up_{\pi\pi}\cdot x}\phi_{\pi\pi}(u),
 \nonumber\\
 \langle {(\pi^+\pi^-)_S}|\bar s (x)  s(0)|0\rangle &=&F_{\pi\pi}^{s}(m_{\pi\pi}^2) B_0
 \int_0^1 du e^{iup_{\pi\pi}\cdot x}\phi_{\pi\pi}^s(u),
 \nonumber \\
 \langle {(\pi^+\pi^-)_S}\bar s(x)\sigma_{\mu\nu}   s(0)|0\rangle &=&
 -F_{\pi\pi}^{s}(m_{\pi\pi}^2)   B_0\frac{1}{6}  (p_{{\pi\pi}\mu} x_\nu -p_{{\pi\pi}\nu} x_\mu) \int_0^1 du e^{i up_{\pi\pi}\cdot x} {\phi_{\pi\pi}^\sigma(u)}.
\end{eqnarray}

The LCDA $\phi_{\pi\pi}$ is twist-2, and the other two  are twist-3. Their normalisations are given as
\begin{eqnarray}
 \int_0^1 du \phi_{\pi\pi}^s(u)=\int_0^1
 du\phi_{\pi\pi}^\sigma(u)=1.
\end{eqnarray}
The    conformal symmetry in QCD~\cite{Braun:2003rp} indicates  that the twist-3 LCDA  have the asymptotic form~\cite{Diehl:1998dk,Polyakov:1998ze,Kivel:1999sd,Diehl:2003ny}:
\begin{eqnarray}
 \phi_{\pi\pi}^s(u)=1, \;\;\; \phi_{\pi\pi}^\sigma(u)=  6u(1-u),
\end{eqnarray}
while the twist-2 LCDA can be expanded in terms of the Gegenbauer moments:
\begin{eqnarray}
 \phi_{\pi\pi}(u)=  6u(1-u) \sum_{n} a_n C_n^{3/2}(2u-1).
\end{eqnarray}
In most cases, contributions from higher Gegenbauer moments are suppressed and thus one may keep the lowest    moment $a_1$.
It is  worthwhile  to stress that these   LCDAs for a two-hadron system  have the same form as the ones for a light scalar $\bar qq$ meson~\cite{Diehl:1998dk,Polyakov:1998ze,Kivel:1999sd,Diehl:2003ny}.  In Ref.~\cite{Cheng:2005nb}, the first Gegenbauer moment for the $f_0(980)$ is calculated as~\footnote{In Ref.~\cite{Cheng:2005nb}, the normalization factor   for a scalar $\bar qq$ meson is $m_{f_0(980)} f_{f_0(980)}$, which is the $F_{\pi\pi}^s B_0$ in this work. The results for the twist-2 Gegenbauer moment in Eqs.~\eqref{eq:a_1_cheng} and \eqref{eq:a_1_pqcd} have been converted to our convention.  }
\begin{eqnarray}
a_1=-1.35.  \label{eq:a_1_cheng}
\end{eqnarray}
while the perturbative QCD analysis of  the $B_s$ decays has used a much smaller value~\cite{Wang:2015uea}:
\begin{eqnarray}
a_1=-0.36. \label{eq:a_1_pqcd}
\end{eqnarray}

\section{Heavy-to-Light Form Factors}
\label{sec:heavy_to_light}

 The $B_s\to \pi^+\pi^-$ transition can be parametrized by three form factors
 \begin{eqnarray}
 \langle (\pi^+\pi^-)_S|\bar s \gamma_\mu\gamma_5 b|\overline B_s
 \rangle  &=&  \frac{-i }{m_{B_s}} \bigg\{ \bigg[P_{\mu}
 -\frac{m_{B_s}^2-m_{\pi\pi}^2}{q^2} q_\mu \bigg] {\cal F}^{B_s\to \pi\pi}_{1}(m_{\pi\pi}^2, q^2)
 +\frac{m_{B_s}^2-m_{\pi\pi}^2}{q^2} q_\mu  {\cal F}^{B_s\to \pi\pi}_{0}(m_{\pi\pi}^2, q^2)  \bigg\},
 \nonumber\\
 \langle (\pi^+\pi^-)_S|\bar s \sigma_{\mu\nu} q^\nu \gamma_5 b|
 \overline B_s \rangle  &=& \frac{{\cal F}^{B_s\to \pi\pi}_T(m_{\pi\pi}^2,
 q^2)}{m_{B_s}(m_{B_s}+m_{\pi\pi})} \bigg[ ({m_{B_s}^2-m_{\pi\pi}^2}) q_\mu - q^2
 P_{\mu}\bigg],
 \label{eq:generalized_form_factors}
\end{eqnarray}
 where  the orbital angular momentum in the $\pi^+\pi^-$ system  is chosen as zero in order to select the S-wave.
The $p_{B_s}$  and $p_{\pi\pi}$ is the momentum for the $B_s$ and the $\pi\pi$ system, respectively.  The momentum transfer is defined as $q=p_{B_s}-p_{\pi\pi}$, and  the $P_\mu$ is defined as $P_{\mu}= p_{B_s}+p_{\pi\pi}$. Here the convention  slightly differs with the ones adopted  in Refs.~\cite{Doring:2013wka,Meissner:2013pba,Meissner:2013hya}.  The $D_s\to \pi^+\pi^-$ form factors can be analogously defined, with the replacement  $m_{B_s}\to m_{D_s}$.

As we have demonstrated in Ref.~\cite{Meissner:2013hya},  the LCSR allows us to express the $B_s\to \pi^+\pi^-$ form factors   in terms of  the $\pi^+\pi^-$ LCDAs~\cite{Mueller:1998fv,Diehl:1998dk,Polyakov:1998ze,Kivel:1999sd,Diehl:2003ny,Hagler:2002nh,Pire:2008xe}.  The   LCSR  factorization formulas  read as~\cite{Meissner:2013hya},
\begin{widetext}
\begin{eqnarray}
  {\cal F}^{B_s\to \pi\pi}_1(m_{\pi\pi}^2, q^2)&=&  N_F
  \bigg\{\int_{u_0}^1\frac{du}{u}{\rm exp}\left[-\frac{m_b^2+u\bar u m_{\pi\pi}^2-\bar uq^2}{uM^2}\right]  \bigg[-\frac{m_b}{B_0}\Phi_{\pi\pi}(u)+u \Phi_{\pi\pi}^s(u)+\frac{1}{3} \Phi_{\pi\pi}^\sigma(u) \nonumber\\
  &&   +\frac{
 m_b^2+q^2-u^2m_{\pi\pi}^2}{uM^2}\frac{\Phi_{\pi\pi}^\sigma(u)}{6}
 \bigg]
 +\exp{\left[-\frac{s_0}{M^2}\right]}\frac{\Phi_{\pi\pi}^\sigma(u_0)}{6}\frac{m_b^2-u_0^2m_{\pi\pi}^2+q^2}
 {m_b^2+u_0^2m_{\pi\pi}^2-q^2}\bigg\},
 \label{eq:fplus}
 \end{eqnarray}
 \begin{eqnarray}
  {\cal F}^{B_s\to \pi\pi}_-(m_{\pi\pi}^2, q^2)&=&  N_F\left\{\int_{u_0}^1\frac{du}{u}{\rm
 exp}\left[-\frac{m_b^2+u\bar u m_{\pi\pi}^2-\bar uq^2}{uM^2}\right]
 \bigg[ \frac{m_b}{B_0}\Phi_{\pi\pi}(u)+(2-u)  \Phi_{\pi\pi}^s(u)\right.
 \nonumber\\
 &&\;\;\;\left. +\frac{1-u}{3u}\Phi_{\pi\pi}^\sigma(u) -\frac{u({m_b^2+q^2-u^2m_{\pi\pi}^2})+2(
 m_b^2-q^2+u^2m_{\pi\pi}^2)}{u^2M^2}\frac{ \Phi_{\pi\pi}^\sigma(u)}{6}
 \bigg]\right.\nonumber\\
 &&\left. -\frac{u_0({m_b^2+q^2-u_0^2m_{\pi\pi}^2})+2(
 m_b^2-q^2+u_0^2m_{\pi\pi}^2) }{u_0(m_b^2+u_0^2m_{\pi\pi}^2-q^2)}
 \exp{\left[-\frac{s_0}{M^2}\right]}\frac{ \Phi_{\pi\pi}^\sigma(u_0)}{6}\right\},
  \label{eq:fminus}
  \end{eqnarray}
 \begin{eqnarray}
    {\cal F}^{B_s\to \pi\pi}_0(m_{\pi\pi}^2, q^2)&=&  {\cal F}^{B_s\to \pi\pi}_1(m_{\pi\pi}^2, q^2) + \frac{q^2}{m_{B_s}^2 -m_{\pi\pi}^2}   {\cal F}^{B_s\to \pi\pi}_-(m_{\pi\pi}^2, q^2),
  \end{eqnarray}
 \begin{eqnarray}
  {\cal F}^{B_s\to \pi\pi}_T(m_{\pi\pi}^2, q^2)&=&2   N_F (m_{B_s}+m_{\pi\pi})
 \bigg\{\int_{u_0}^1\frac{du}{u} {\rm exp}\left[-\frac{(m_b^2-\bar uq^2+u\bar
 um_{\pi\pi}^2)}{uM^2}\right]
 \left[-\frac{\Phi_{\pi\pi}(u)}{2B_0}+m_b\frac{ \Phi_{\pi\pi}^\sigma(u)}{6uM^2}\right] \nonumber\\
 &&   +m_b\frac{ \Phi_{\pi\pi}^\sigma(u_0)}{6}
   \frac{\exp[-s_0/M^2]}{m_b^2-q^2+u_0^2m_{\pi\pi}^2}\bigg\},
   \label{eq:ftensor}
\end{eqnarray}\end{widetext}
where
\begin{eqnarray}
 N_F &=&B_0  F^s_{\pi\pi}(m_{\pi\pi}^2) \frac{m_b+m_s}{2m_{B_s}f_{B_s}} {\rm
 exp}\left[\frac{m_{B_s}^2}{M^2}\right],\nonumber\\
 u_0&=&\frac{m_{\pi\pi}^2+q^2-s_0+\sqrt{(m_{\pi\pi}^2+q^2-s_0)^2+4m_{\pi\pi}^2(m_b^2-q^2)}}{2m_{\pi\pi}^2}~.
 \label{eq:u0}
\end{eqnarray}
In order to derive the above equations,
the  Borel transformation of   hadronic and of   QCD  expressions of
correlation functions has been   carried out,   defined as:
\begin{eqnarray} {\cal B} [{\cal
F}(Q^2)]={\rm lim}_{Q^2 \to \infty, \; n \to \infty, \; {Q^2 \over
n}=M^2} {1 \over (n-1)!} (-Q^2)^n \left({d \over dQ^2} \right)^n
{\cal F}(Q^2) \; ,
\label{tborel} \end{eqnarray}
where ${\cal F}$  is a  function of $Q^2=-q^2$ and $M^2$ is  the Borel parameter.  The explicit form is:
 \begin{eqnarray}
 {\cal B} \left[ { 1 \over (s+Q^2)^n }
\right]={\exp(-s/M^2) \over (M^2)^n\ (n-1)!} \; .
\label{bor} \end{eqnarray}
This operation improves the convergence of the OPE series  by
factorials of $n$ and, for suitably chosen values of $M^2$,
enhances the contribution of the low-lying states to the correlation function.

For convenience we can define the normalized form factor:
\begin{eqnarray}
 {\cal F}_i(m_{\pi\pi}^2, q^2)&=& B_0 F_{\pi\pi}^s(m_{\pi\pi}^2) \overline F_i(q^2),
\end{eqnarray}
where the  $m_{\pi\pi}$ and $q^2$ dependence   has been    factorized  into the $F_{\pi\pi}^s(m_{\pi\pi}^2)$ and $\overline F_i(q^2)$, respectively. This approximation  is   justified by the Watson-Madigal theorem~\cite{Watson:1952ji,Migdal:1955}.  As a reference point, we will choose the $m_{\pi\pi}=m_{f_0(980)}$ to explore   the functions  $\overline F_i(q^2)$.

\begin{figure}\begin{center}
\includegraphics[scale=0.5]{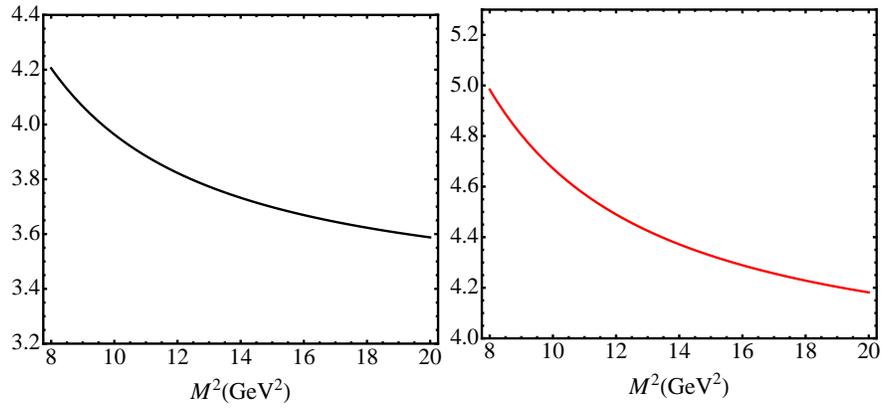}
\caption{At  the maximal recoil    $q^2=0$, the dependence of  $\overline F_1(q^2=0)=\overline F_0(q^2=0)$  (left panel) and  $\overline F_T(q^2=0)$ (right panel) on the Borel parameter $M^2$.  The final results are obtained requiring stability against variations of $M^2$.   } \label{fig:F10T_M2}
\end{center}
\end{figure}

In the numerical analysis, we use~\cite{Aoki:2013ldr,Agashe:2014kda}
\begin{eqnarray}
f_{B_s}= (224\pm 5) {\rm MeV},\;\;\; s_0=(34\pm 2) {\rm GeV}^2,  \;\;\; B_0= (1.7\pm0.2) {\rm GeV}.
\end{eqnarray}
With these  numerical inputs,   the sum rules
(\ref{eq:fplus})-(\ref{eq:ftensor}) provide us with  the functions ${\overline F}_i(q^2)$ for
each value of $q^2$ as a  function of the Borel parameter. In Fig.~\ref{fig:F10T_M2}, at $q^2=0$ we show  the dependence on  $M^2$ with $a_1=-0.6$. The results are obtained requiring stability against variations of
$M^2$.  As demonstrated in this figure, the form factors become  stable  when $M^2>12{\rm GeV}^2$.  The situations with different $a_1$ and $q^2$ values are similar and thus we can choose $M^2= (16\pm2){\rm GeV}^2$.

\begin{figure}\begin{center}
\includegraphics[scale=0.4]{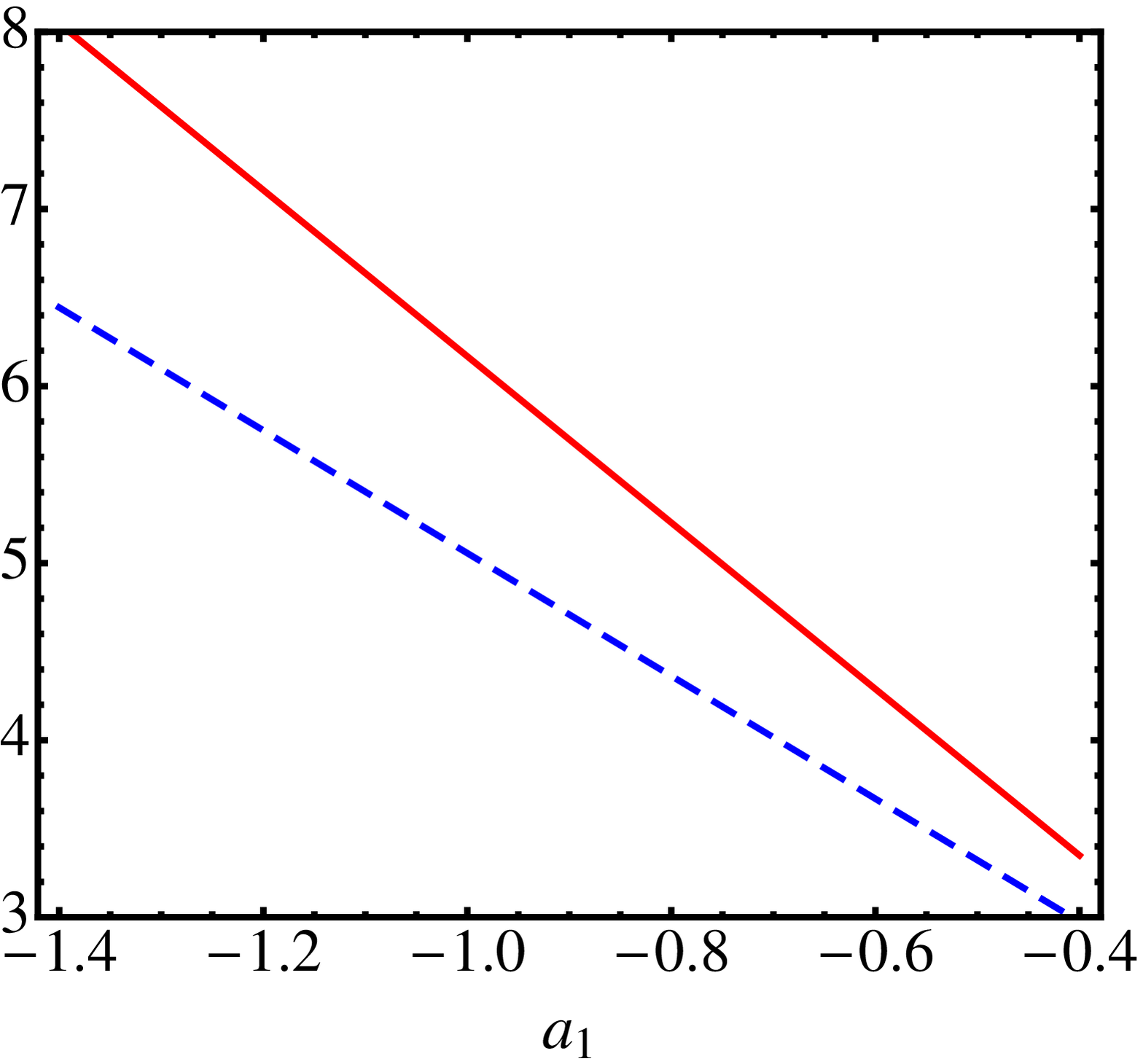} \hspace{1.cm}
\includegraphics[scale=0.4]{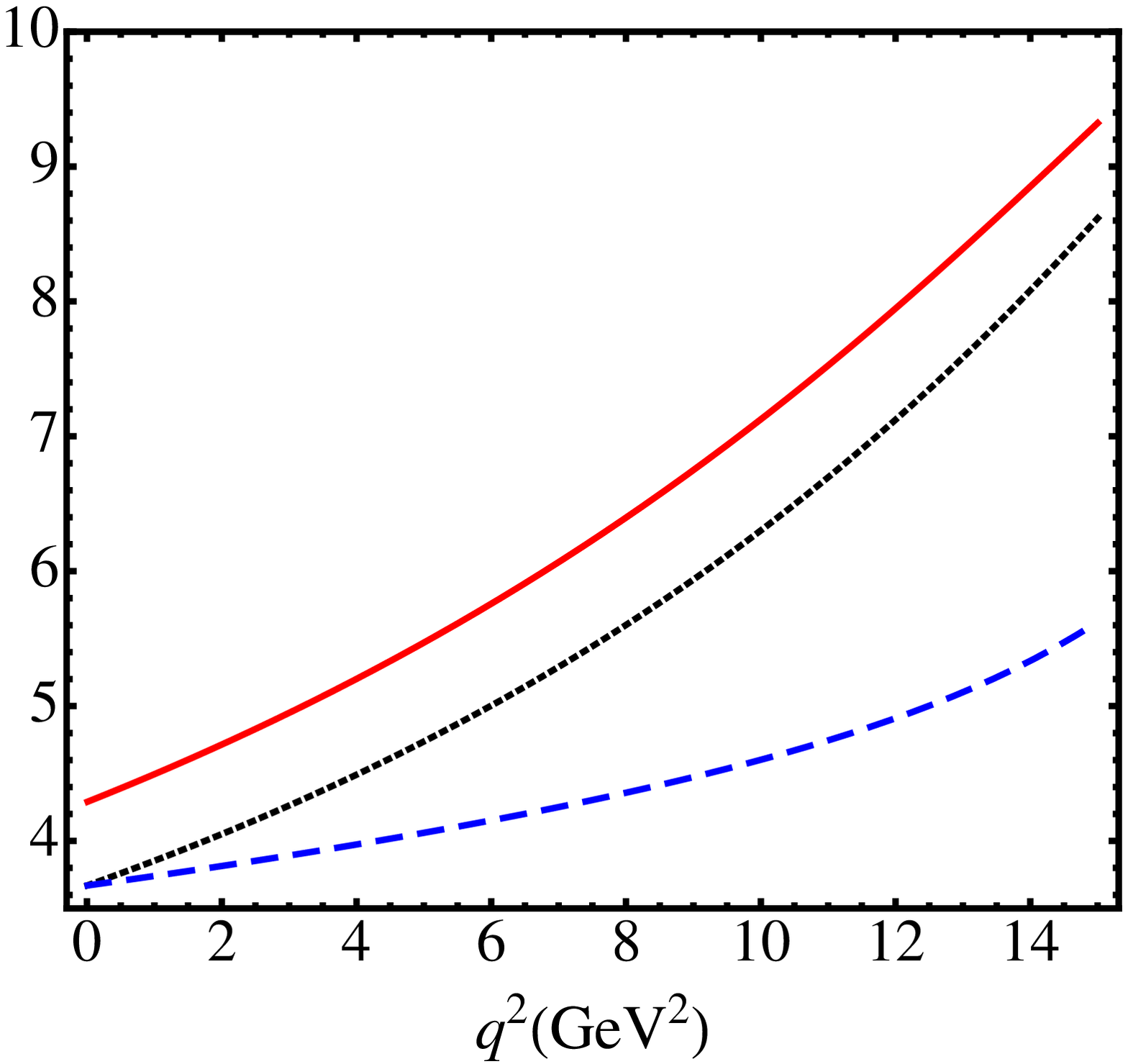}
\caption{At the maximal recoil $q^2=0$,  the dependence of $\overline F_{1,0}$ , and   $\overline F_T$ on the Gegenbauer moment $a_1$ is shown in the left panel. Dashed and solid curves correspond to  $\overline F_1(q^2=0) = \overline F_0(q^2=0)$, and  $\overline F_T(q^2=0)$, respectively.  In the right panel,   the $q^2$ dependence is given  with   $a_1=-0.6$. Solid, dotted and dashed lines denote the $\overline F_T(q^2)$, $\overline F_1(q^2)$ and $\overline F_0(q^2)$, respectively.  } \label{fig:F10T_a1_q2}
\end{center}
\end{figure}

Since the form factors are sensitive to the Gegenbauer moment $a_1$ of the twist-2 LCDA,  in the left panel of Fig.~\ref{fig:F10T_a1_q2} we show this  dependence in the range $a_1=(-1.4, -0.4)$ at the maximal recoil $q^2=0$.   We will show later the value $a_1=-0.6$  can describe well the data on both the $B_s\to\pi^+\pi^-\ell^+\ell^-$ and $D_s\to \pi^+\pi^-\ell\nu$.

The LCSR is applicable in the hard-scattering region.
To access the momentum distribution  in the full kinematics region, we will adopt the  following parametrization
\begin{eqnarray}
\overline  F_i(q^2)= \frac{\overline F_i(0)}{1-a_i q^2/m_{B_s}^2 +b_i (q^2/m_{B_s}^2)^2}, \label{eq:q2-dependence}
\end{eqnarray}
where $i=1,0,T$.  The   parameters can be fitted in the region $q^2< 5{\rm GeV}^2$ for the $B_s$ transition, and the results are collected in Tab.~\ref{table:q2_ff}.

\begin{table}
\caption{Fitted parameters of the $B_s/D_s \to \pi^+\pi^-$  form factors derived  by LCSR.   }
 \label{table:q2_ff}
\begin{tabular}{ c c c c |c c c c c c}
\hline\hline $B_s\to \pi^+\pi^-$ & $\overline F_i(q^2=0)$  & $a_i$ & $b_i$  &$D_s\to \pi^+\pi^-$ & $\overline F_i(q^2=0)$  & $a_i$ & $b_i$
\\\hline
 $\overline F_1 $   &   $3.66$   &   $1.39$  &   $0.54$ & $\overline F_1 $   &   $2.45$   &   $0.82$  &   $0.20$ \\
 $\overline F_0 $   &   3.66 &    0.54& $-0.08$ & $\overline F_0 $   &   $2.45$   &   $0.39$  &   $-0.15$ \\
 $\overline F_T $   &   4.29 & 1.33 & 0.54\\
\hline\hline
\end{tabular}
\end{table}

\begin{figure}\begin{center}
\includegraphics[scale=0.4]{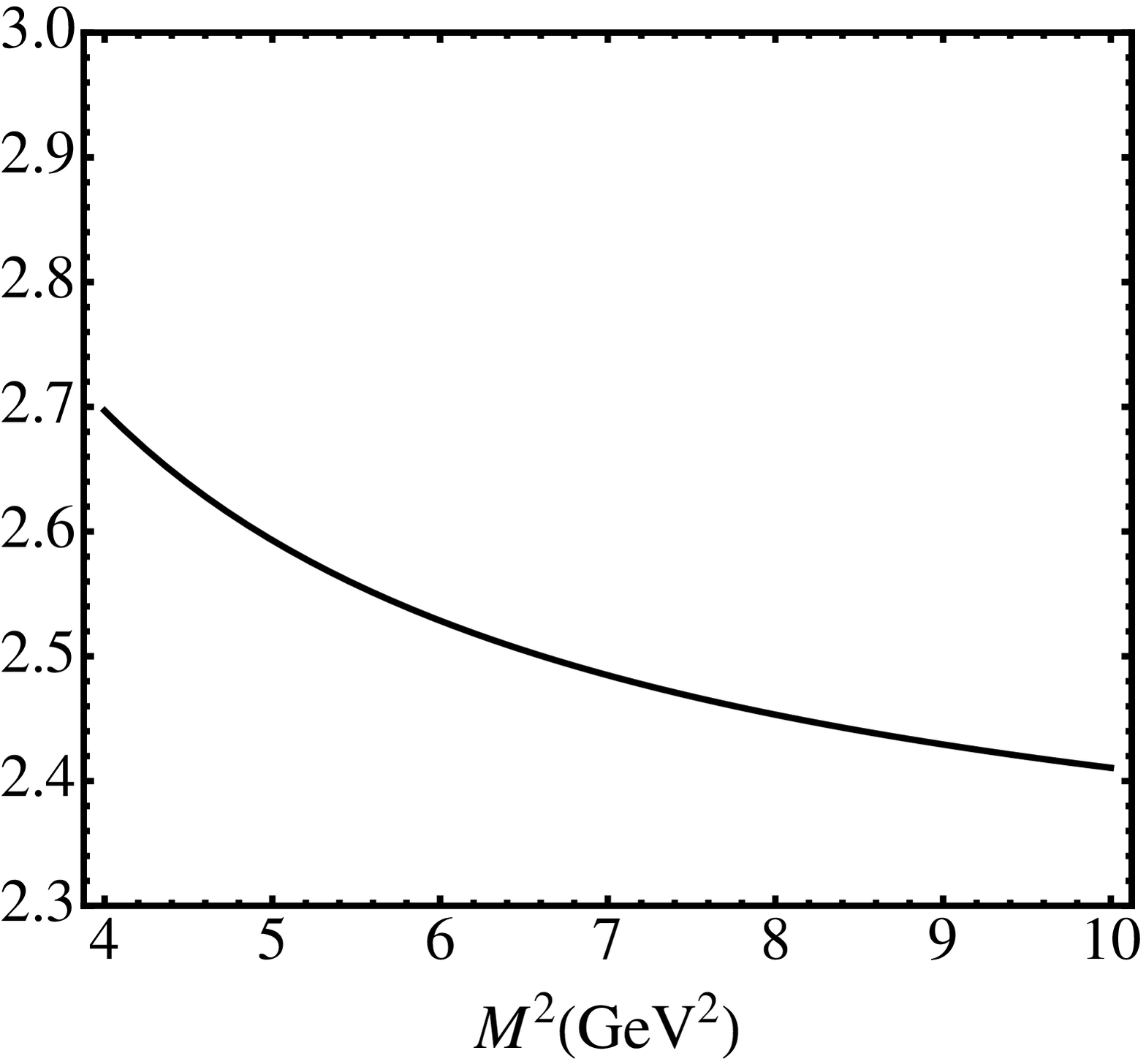}
\hspace{1.cm}
\includegraphics[scale=0.4]{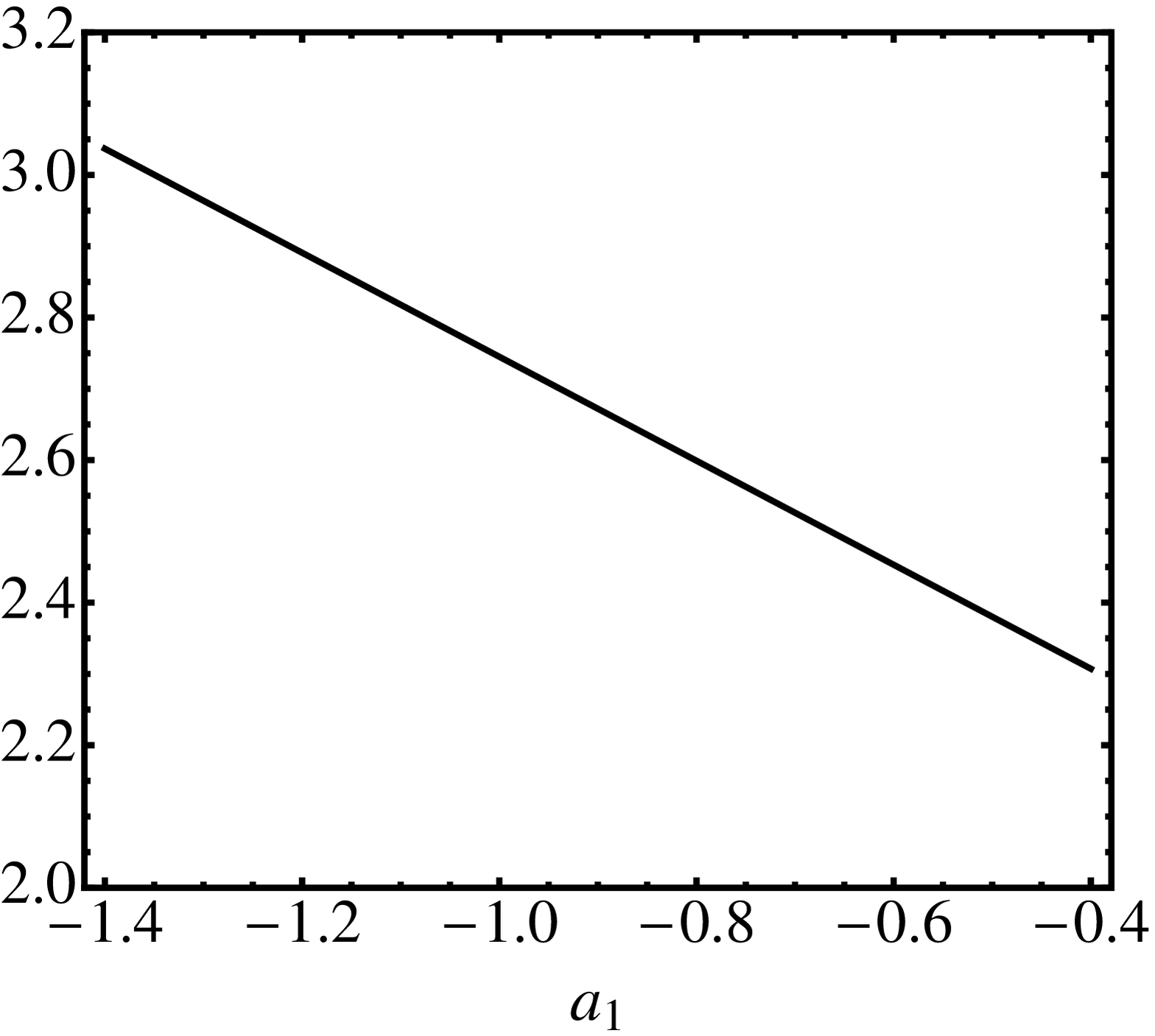}
\caption{The functions $\overline F_i(0)$ for the $D_s\to \pi^+\pi^-$: the dependence on $M^2$(Gegenbauer moment $a_1$)  in the left  (right) panel.  } \label{fig:F10T_M2_a1_Ds}
\end{center}
\end{figure}

For  the $D_s\to \pi^+\pi^-$ transition,  we use~\cite{Agashe:2014kda}
\begin{eqnarray}
 s_0=(6.5\pm 1) {\rm GeV}^2, f_{D_s}= (257.5\pm 4.6) {\rm MeV},\;\;\; m_c=1.4 {\rm GeV}.
\end{eqnarray}
The results for the $D_s\to \pi^+\pi^-$ form factors are given in Fig.~\ref{fig:F10T_M2_a1_Ds}.  From the left panel, we can see the results are stable when $M^2> 6{\rm GeV}^2$, and we will use $M^2=(8\pm1){\rm GeV}^2$.  The dependence on the first Gegenbauer moment $a_1$ is less severe compared to the $B_s\to \pi^+\pi^-$ case, as shown in the right panel of Fig.~\ref{fig:F10T_M2_a1_Ds}.  In the $D_s$ mode,   the twist-2 contributions in the regions with $x>1/2$ and $x<1/2$ cancel with each other. This fact has been explored in the study of $D_s\to f_0(980)$ transition~\cite{Colangelo:2010bg}.
Since the energy release in the $D_s$ transition is small, we have used the $-5{\rm GeV}^2<q^2<0$ to fit the $q^2$-dependent parameters in Eq.\eqref{eq:q2-dependence}.

\section{Phenomenological Results}
\label{sec:phono}

\subsection{$B_s\to \pi^+\pi^- \ell^+\ell^-$}

We   proceed  with the analysis of  $B_s\to \pi^+\pi^-\ell^+\ell^-$, whose  decay amplitude is governed by the  effective Hamiltonian~\cite{Buchalla:1995vs}
\begin{eqnarray}
 {\cal
 H}_{\rm{eff}}=
 -\frac{G_F}{\sqrt{2}}V_{tb}V^*_{ts}\sum_{i=1}^{10}C_i(\mu)O_i(\mu).
 \nonumber\label{eq:Hamiltonian}
 \end{eqnarray}
The $O_i$ is a four-quark or a  magnetic-moment operator,  and the $C_i(\mu)$ is its    Wilson coefficient. The explicit forms  can be found in Ref.~\cite{Buchalla:1995vs}.
$G_F$ is the Fermi constant, and $V_{tb}=0.99914\pm0.00005$ and
$V_{ts}=-0.0405^{+0.011}_{-0.012}$~\cite{Agashe:2014kda} are the CKM matrix elements.   The bottom and strange quark masses are $m_b=(4.66\pm 0.03)$GeV and
$m_s=(0.095\pm0.005)$GeV~\cite{Agashe:2014kda}.


In general, various partial-waves of two-hadron $M_1M_2$ state   contribute to a generic   $B\to M_1M_2\ell^+\ell^-$  process and the  differential decay width has been derived  using the helicity amplitude in Refs.~\cite{Lu:2011jm,Dey:2015rqa,Gratrex:2015hna}.  In the $B_s\to \pi^+\pi^-\mu^+\mu^-$,    the S-wave contribution dominate with the angular distribution:
\begin{eqnarray}
 \frac{d^3\Gamma(B_s\to \pi^+\pi^-\mu^+\mu^-)}{dm_{\pi\pi}^2dq^2  d\cos\theta_{\ell} }
 &=& \frac{3}{8}\Big[J_1^c   + J_2^c
 \cos(2\theta_{\ell})  \Big],
\end{eqnarray}
where the   coefficients  are
\begin{eqnarray}
J_1^c&=&     |{\cal A}^0_{L0}|^2+|{\cal A}^0_{R0}|^2
 +8  \hat m_{\ell}^2  | {\cal A}^0_{L0}{\cal A}^{0*}_{R0} | \cos(\delta_{L0}^0 -\delta_{R0}^0)
 +4 \hat m_{\ell}^2  |{\cal A}_t^0|^2  ,  \\
 J_2^c  &=& -\beta_{2\ell}^2   \bigg\{   |{\cal A}^0_{L0}|^2+|{\cal A}^0_{R0}|^2    \bigg\} .
  \label{eq:simplified_angularCoefficients_S-wave}
\end{eqnarray}
In the above equations, $\beta_{2\ell}=
\sqrt{1-4\hat m_\ell^2},
\hat m_\ell= m_\ell/\sqrt{q^2}$, and
$\theta_{\ell}$  is the polar angle  between  the $B_s$ and the  $\mu^-$   moving direction  in the  lepton pair rest-frame.  The $\delta_{L0}^0$ and $\delta_{R0}^0$ are phases of the   helicity amplitudes
\begin{eqnarray}
 {\cal A}_{L/R,0}^0&=&N_1^{2\ell}\sqrt{N_2^{2\ell}} i\frac{1}{m_{B_s}}\Bigg[ (C_9\mp C_{10}) \frac{\sqrt {\lambda}}{\sqrt{ q^2}} {\cal F}_1^{B_s\to\pi\pi}(q^2) +2(C_{7L}-C_{7R})  \frac{\sqrt {\lambda }m_b}{\sqrt {q^2}(m_B+m_{\pi\pi})}{\cal F}^{B_s\to\pi\pi}_T(q^2)  \Bigg],\nonumber\\
  {\cal A}_{L/R,t}^0&=&N_1^{2\ell}\sqrt{N_2^{2\ell}} i\frac{1}{m_{B_s}}\Bigg[ (C_9\mp C_{10}) \frac{m_{B_s}^2-m_{\pi\pi}^2}{\sqrt {q^2}} {\cal F}^{B_s\to\pi\pi}_0(q^2) \Bigg],\label{eq:helicity_ALR0}\\
{\cal A}_{t}^0&=&{\cal A}_{R,t}^0-{\cal A}_{L,t}^0=  2N_1^{2\ell} \sqrt{N_2^{2\ell}}C_{10}   i\frac{1}{m_{B_s}}\Bigg[  \frac{m_{B_s}^2-m_{\pi\pi}^2}{\sqrt {q^2}} {\cal F}^{B_s\to\pi\pi}_0(q^2) \Bigg].
\end{eqnarray}
where
\begin{eqnarray}
N_1^{2\ell}&=& \frac{ G_F} {4\sqrt 2} \frac{\alpha_{\rm em}}{\pi} V_{tb}V_{ts}^*\\
N_2^{2\ell}&=& \frac{1}{16\pi^2}\sqrt{1-4m_{\pi}^2/m_{\pi\pi}^2} \times  {\frac{8}{3}} \frac{\sqrt {\lambda}
{q^2}\beta_{2\ell}}{256\pi^3 m_{B_s}^3}.
\end{eqnarray}
The   K\"allen function $\lambda$ is related to the $\pi^+\pi^-$ momentum in the $B_s$ rest-frame:
\begin{eqnarray}
\lambda\equiv\lambda(m^2_{B_s},m^2_{\pi^+\pi^-},
q^2),\;\;\;\;
 \lambda (a,b,c) = a^2+ b^2 +c^2- 2(ab+bc+ca)~.
\end{eqnarray}

\begin{figure}
\begin{center}
\includegraphics[scale=0.5]{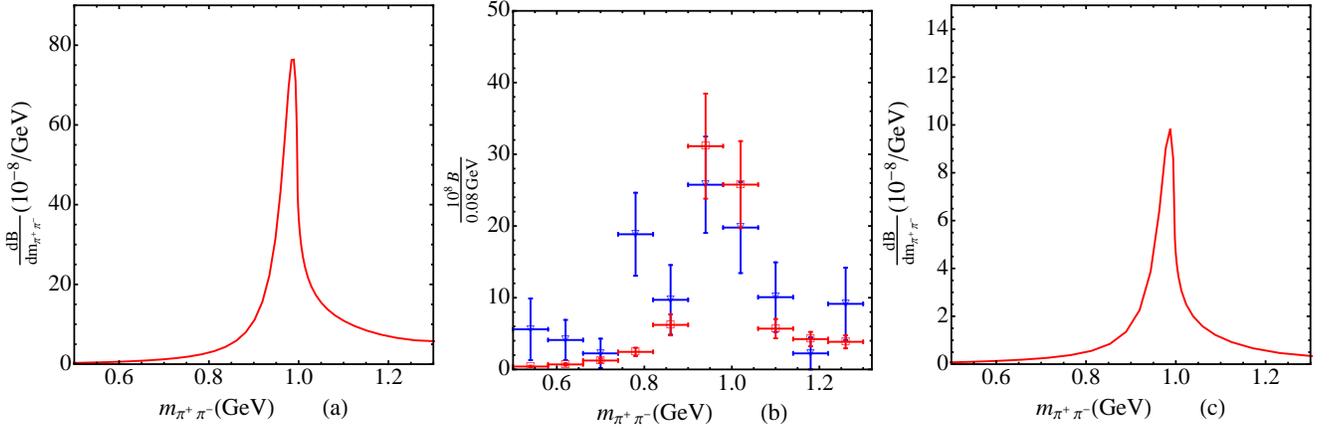}
\caption{Differential branching ratios $d{\cal B}/dm_{\pi\pi}$  for the $B_s\to \pi^+\pi^-\mu^+\mu^-$ in panel (a) and (b), and $B_s\to \pi^+\pi^-\tau^+\tau^-$ in panel (c).  In panel (b),   experimental data (with triangle markers)  has been normalized to the central value of the branching fraction: ${\cal B}(B_s^0\to \pi^+\pi^-\mu^+\mu^-)=(8.6\pm 1.5\pm 0.7\pm 0.7)\times 10^{-8}$~\cite{Aaij:2014lba}, and  theoretical results are shown with square markers. } \label{fig:dBRdM2h}
\end{center}
\end{figure}

In Fig.~\ref{fig:dBRdM2h},  results for differential branching fractions  $d{\cal B}/dm_{\pi\pi}$  for the $B_s\to \pi^+\pi^-\mu^+\mu^-$  are given   in  panel (a) and (b), and  the ones for $B_s\to \pi^+\pi^-\tau^+\tau^-$ are given in panel (c).   The result in panel (a)  clearly shows the peak corresponding to the $f_0(980)$.  In order to compare with the experimental data~\cite{Aaij:2014lba}, we also give the binned results in  panel (b) in Fig.~\ref{fig:dBRdM2h} from $0.5$ GeV to $1.3$ GeV with square markers.  Dominant theoretical errors arise from the $B_0=(1.7\pm0.2){\rm GeV}$. The experimental data (with triangle markers)  has been normalized to the central value in Eq.~\eqref{eq:Bspipidata}.   The comparison in this panel shows an overall  agreement  between  our theoretical predictions  and the experimental data.
Integrating out the $m_{\pi\pi}$ from $0.5$ GeV to $1.3$ GeV, we have the branching fraction:
 \begin{eqnarray}
{\cal B}(B_s\to   \pi^+\pi^-  \mu^+\mu^-) = (6.9\pm 1.6)\times 10^{-8}, \label{eq:theory_Bs_f0_mumu}
 \end{eqnarray}
 which is also  consistent with the data in  Eq.~\eqref{eq:Bs_f0_mumu_data}.

\begin{figure}\begin{center}
\includegraphics[scale=0.5]{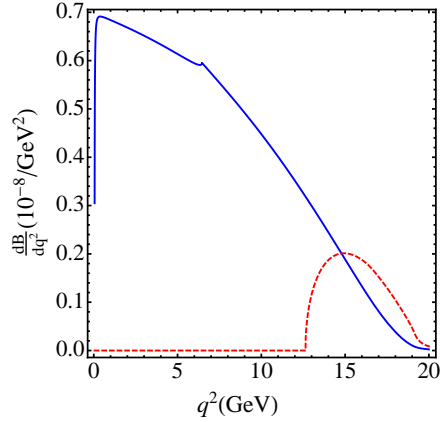}
\caption{The differential branching ratio $d{\cal B}/dq^2$  for the $B_s\to \pi^+\pi^-\mu^+\mu^-$ (solid curve) and  $B_s\to \pi^+\pi^-\tau^+\tau^-$(dashed curve) is given in unit of $10^{-8}/{\rm GeV}^2$.  } \label{fig:dBRdq2}
\end{center}
\end{figure}

In Fig.~\ref{fig:dBRdq2}, we  predict  the differential  distribution $d{\cal B}/dq^2 $ (in unit of $10^{-8}/{\rm GeV}^2$) for  the $B_s\to \pi^+\pi^-\mu^+\mu^-$ (solid curve) and   for  the $B_s\to \pi^+\pi^-\tau^+\tau^-$ (dashed curve).
Results for the  integrated branching fractions of $B_s\to \pi^+\pi^-\tau^+\tau^-$  are  predicted  as:
 \begin{eqnarray}
{\cal B}(B_s\to   \pi^+\pi^-  \tau^+\tau^-) = (8.8\pm 2.1)\times 10^{-9}, \label{eq:theory_Bs_f0_tautau}
 \end{eqnarray}
 where $0.5{\rm GeV}<m_{\pi\pi}<1.3$ GeV is assumed.
Our theoretical  results could be examined at the future experimental facilities including the LHCb detector~\cite{Bediaga:2012py} and    the Super-B factory at the KEK~\cite{Aushev:2010bq}.

\subsection{$B_s\to \pi^+\pi^-\nu\bar\nu$}

The  $b \to s \nu \bar \nu$ effective Hamiltonian  is given by
\begin{equation}
{\cal H}_{b \to s\nu \bar \nu}= {G_F \over \sqrt{2}} {\alpha_{em} \over 2 \pi
\sin^2(\theta_W)} V_{tb} V_{ts}^* \eta_X X(x_t) \, O_L  \equiv C_L
O_L \,\,, \label{eq:hamilnu}
\end{equation}
which involves the four-fermion operator
\begin{equation}
O_L = [{\bar s}\gamma^\mu (1-\gamma_5) b ][{\bar \nu}\gamma_\mu
(1-\gamma_5) \nu] \,\, .\label{opnu}
\end{equation}
Here $\theta_W$  is the Weinberg angle; the  function $X(x_t)$
($x_t= { m_t^2 / m_W^2}$,  with $m_t$  the top
quark mass and $m_W$ the $W$ mass) has been computed in Refs.~\cite{Inami:1980fz,Buchalla:1995vs}, and   the QCD factor $\eta_X$ is found  close to one
\cite{Buchalla:1993bv,Buchalla:1998ba,Misiak:1999yg}.

\begin{figure}\begin{center}
\includegraphics[scale=0.5]{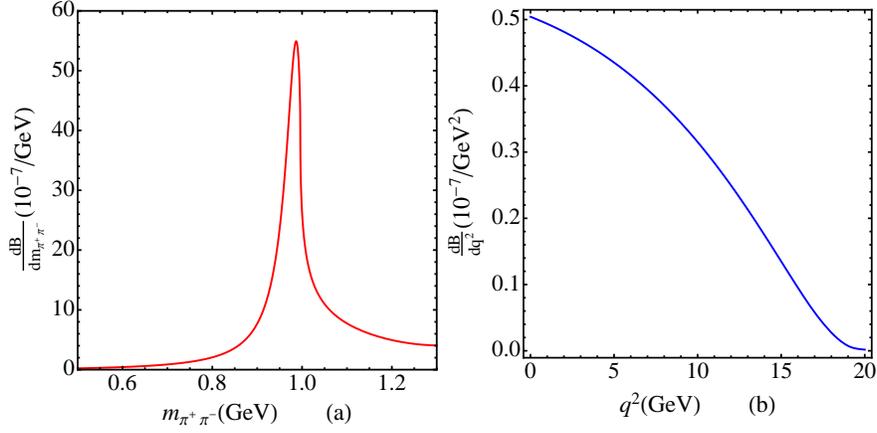}
\caption{The differential branching ratios for the $B_s\to \pi^+\pi^-\nu\bar\nu$: the left panel  for the $d{\cal B}/dm_{\pi\pi}$, and the right one for the $d{\cal B}/dq^2$.  } \label{fig:dBR2nu}
\end{center}
\end{figure}

With  the above  Hamiltonian,  we obtain the differential decay width
\begin{eqnarray}
 \frac{d^2\Gamma(\bar B_s\to (\pi^+\pi^-)\nu\bar\nu)}{dq^2dm_{\pi\pi}^2}&=&3 \times |A_0^0|^2,
\end{eqnarray}
where the factor $3$ arises from three species of neutrinos.
The helicity amplitude in this case is
\begin{eqnarray}
 A_0^0 &=& C_L\sqrt{ N_2^{\nu}} i\frac{1}{m_{B_s}} \Bigg[   \frac{\sqrt {\lambda}}{\sqrt{ q^2}}{\cal F}_1^{B_s\to \pi\pi}(m_{\pi\pi}^2, q^2)  \Bigg], \nonumber\\
N_2^{2\nu}&=& \frac{1}{16\pi^2}\sqrt{1-4m_{\pi}^2/m_{\pi\pi}^2} \times  {\frac{8}{3}} \frac{\sqrt {\lambda}
{q^2}}{256\pi^3 m_{B_s}^3}.
\end{eqnarray}
We give our predictions for the differential distributions for the $B_s\to \pi^+\pi^-\nu\bar\nu$  in Fig.~\ref{fig:dBR2nu}: the left panel  for the $d{\cal B}/dm_{\pi\pi}$, and the right one for the $d{\cal B}/dq^2$. The integrated branching fraction in the range $0.5{\rm GeV}<m_{\pi\pi}< 1.3{\rm GeV}$  is predicted as:
\begin{eqnarray}
 {\cal B}(B_s\to \pi^+\pi^-\nu\bar\nu)= (4.9\pm 1.2)\times 10^{-7}.
\end{eqnarray}
There is  a large chance  measure this branching ratio at the Super-B factory at KEK~\cite{Aushev:2010bq}.

 \subsection{$D_s\to \pi^+\pi^- \ell \nu$}

The effective Hamiltonian for $c\to s\ell \nu$ transition is given as
\begin{eqnarray}
 {\cal H}_{c\to s\ell\nu} = N_{1}^{\ell}   [\bar s \gamma_\mu(1-\gamma_5) c]
[ \bar \nu \gamma^\mu(1-\gamma_5) \ell] +h.c.,
\end{eqnarray}
with
\begin{eqnarray}
 N_{1}^{\ell} = \frac{G_F}{\sqrt 2} V_{cs}.
\end{eqnarray}

The differential decay width  for $D_s\to   \pi^+\pi^-\ell   \nu_{\ell}$  can be expressed  as
\begin{eqnarray}
 \frac{d^3\Gamma}{dm_{K\pi}^2dq^2  d\cos\theta_l }
 &=& \frac{3}{8}\Big[I_1(q^2, m_{K\pi}^2)   +I_2 (q^2, m_{K\pi}^2)
 \cos(2\theta_l) +I_6 \cos(\theta_l) \Big],
\end{eqnarray}
with  the $I_i$ having the form:
\begin{eqnarray}
 I_1(q^2, m_{\pi\pi}^2)   &=&  \left[(1+\hat m_l^2) |A^0_{0}|^2
 +2 \hat m_l^2  |A_t^0|^2\right],\nonumber\\
 I_2(q^2, m_{\pi\pi}^2)    &=& -\beta_l     |A^0_{0}|^2,\nonumber\\
 I_6 (q^2, m_{\pi\pi}^2) &=& 4\hat m_l^2 {\rm Re}[A^0_{0} A_t^{0*}].
 \label{eq:simplified_angularCoefficients_Ds_lnu}
\end{eqnarray}
Using the $D_s\to\pi^+\pi^-$ form factors, the  matrix element  for $D_s$ decays
into the S-wave $\pi\pi$  final state  is given as
\begin{eqnarray}
 A_0^0 &=&N_1^{\ell} \sqrt{ N_2^{\ell}} i\frac{1}{m_{D_s}} \Bigg[   \frac{\sqrt {\lambda}}{\sqrt{ q^2}}{\cal F}_1^{D_s\to \pi\pi}(m_{\pi\pi}^2, q^2)  \Bigg],\nonumber\\
 A_t^0 &=&N_1^{\ell}\sqrt{ N_2^{\ell}} i \frac{1}{m_{D_s}}\Bigg[ \frac{m_{D_s}^2-m_{\pi\pi }^2}{\sqrt {q^2}} {\cal F}_0^{D_s\to \pi\pi}(m_{\pi\pi}^2, q^2)\Bigg] ,\label{eq:S-waveKpi-formula}
\end{eqnarray}
where
\begin{eqnarray}
N_2^{\ell}= \frac{1}{16\pi^2}\sqrt{1-4m_{\pi}^2/m_{\pi\pi}^2} \times  {\frac{8}{3}} \frac{\sqrt {\lambda}
{q^2}\beta_{\ell}}{256\pi^3 m_{D_s}^3}.
\end{eqnarray}

\begin{figure}\begin{center}
\includegraphics[scale=0.5]{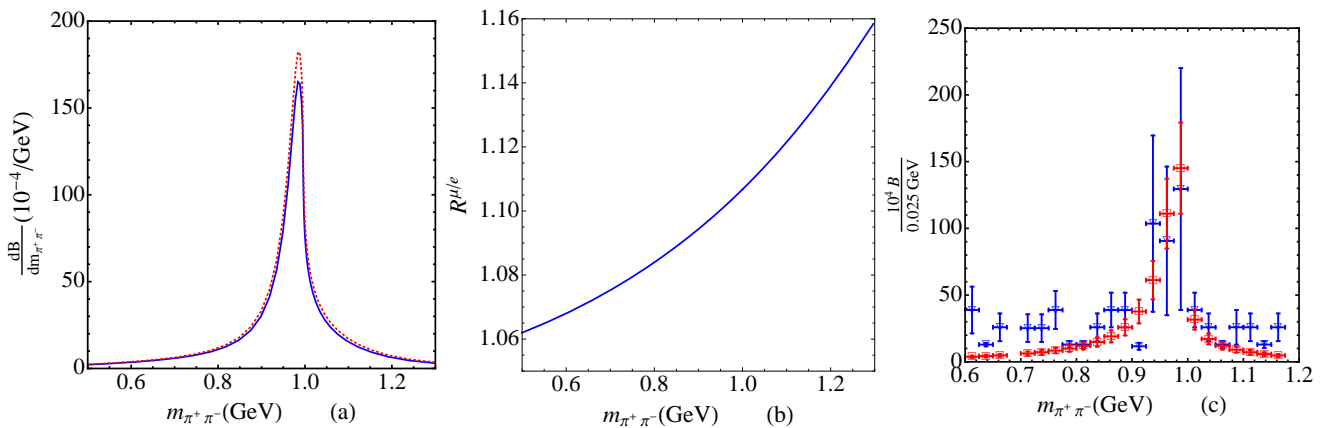}
\caption{The differential branching ratios for the $D_s\to \pi^+\pi^-\ell\nu$. The first  panel corresponds to  the $d{\cal B}/dm_{\pi\pi}$, in which the dotted and solid curves correspond to $D_s\to \pi^+\pi^-\mu\nu$ and $D_s\to \pi^+\pi^-e\nu$ respectively.   The second  panel shows the ratio $R^{\mu/e} (m_{\pi\pi}^2)$ defined in Eq.~\eqref{eq:RKtauOvermuNorm}.  A comparison with the experimental data~\cite{Yelton:2009aa,Ecklund:2009aa} is given in panel (c).   } \label{fig:dBR_Ds_pipi}
\end{center}
\end{figure}

As discussed in Ref.~\cite{Meissner:2013pba}, one can   explore a number of the $q^2$-dependent ratios and in particular the lepton flavor dependent ratio:
\begin{eqnarray}
{\cal R}^{\mu/e}(m_{\pi\pi}^2, q^2) = \frac{ {d^2\Gamma(D_s\to \pi^+\pi^- \mu^+   \nu_{\mu})}/{dq^2dm_{\pi\pi}^2 }}{ {d^2\Gamma(D_s\to \pi^+\pi^-   e^+ \nu_{e})}/{dq^2dm_{\pi\pi}^2 }}, \label{eq:RKtauOvermu}
\end{eqnarray}
and the  integrated form over $q^2$:
\begin{eqnarray}
 R^{\mu/e} (m_{\pi\pi}^2)=  \frac{ {d\Gamma(D_s\to \pi^+\pi^- \mu^+   \nu_{\mu})}/{dm_{\pi\pi}^2 }}{ {d\Gamma(D_s\to \pi^+\pi^-   e^+ \nu_{e})}/{dm_{\pi\pi}^2 }}. \label{eq:RKtauOvermuNorm}
\end{eqnarray}
Our results are given in Fig.~\ref{fig:dBR_Ds_pipi}. The first  panel corresponds to  the $d{\cal B}/dm_{\pi\pi}$, in which the dotted and solid curves denote  to  the $D_s\to \pi^+\pi^-\mu\nu$ and $D_s\to \pi^+\pi^-e\nu$,  respectively.
One different  behavior in the differential  branching ratio  with $B_s\to\pi^+\pi^-\ell^+\ell^-$ in the $m_{\pi\pi}$ distributions is the suppression in the large $m_{\pi\pi}$ region.  The second  panel shows the ratio $R^{\mu/e} (m_{\pi\pi}^2)$ defined in Eq.~\eqref{eq:RKtauOvermuNorm}.  A comparison with the experimental data on the differential   branching fraction ~\cite{Yelton:2009aa,Ecklund:2009aa} is given in panel (c), where we can also find the agreement.

The integrated branching fractions  are predicted  as
\begin{eqnarray}
 {\cal B}( D_s\to \pi^+\pi^-e^+\nu)&=& (1.52\pm0.36)\times 10^{-3}, \\
 {\cal B}( D_s\to \pi^+\pi^-\mu^+\nu)&=& (1.68\pm0.39)\times 10^{-3},
\end{eqnarray}
where $0.5$GeV$<m_{\pi\pi}<1.3 $GeV has been adopted in the integration.  Again the errors come from the QCD condensate parameter $B_0$.   Theoretical results  are in good agreement with the CLEO results in Eqs.~(\ref{eq:Ds_f0_pipi_CLEO_c},\ref{eq:Ds_f0_pipi_CLEO_c_2})~\cite{Yelton:2009aa,Ecklund:2009aa,Hietala:2015jqa}. We expect experimental errors will be greatly reduced since in future the BES-III collaboration will collect about $2fb^{-1}$ data in $e^+e^-$ collision  at the energy around $4.17$GeV which will be used  to study semileptonic and nonleptonic $D_s$ decays~\cite{Asner:2008nq}.

\section{Conclusions}
\label{sec:conclusion}

Rare $B$ decays have played an important role in testing the SM, and hunting for the NP.  In recent years, a lot of experimental progresses have been made on the $B\to K^*\ell^+\ell^-$, and remarkably  the LHCb collaboration has found a $3.7\sigma$  deviation from the SM for the ratio  $P_5'$.   This observable $P_5'$ is believed almost independent on the hadronic uncertainties.

The analysis of $B\to V\ell^+\ell^-$, more appropriately $B\to M_1M_2\ell^+\ell^-$, requests not only the knowledge on the $m_b$ expansion but also the $M_1M_2$ final state interactions.
In this work,  we have studied  the $B_s^0\to \pi^+\pi^-\ell^+\ell^-$,  $B_s^0\to \pi^+\pi^-\nu\bar\nu$ and  $D_s^+\to \pi^+\pi^-\ell^+ \nu$ decay in the kinematics  region where the $\pi^+\pi^-$ system  has a invariant mass in the range $0.5$-$1.3$ GeV. These processes are dominated by the S-wave contributions  and thus they are valuable towards the determination of the S-wave $\pi^+\pi^-$ light-cone distribution amplitudes which are normalized to scalar  form factors. We have compared the results for scalar form factors calculated in  unitarized $\chi$PT and the ones  extracted from the data on the $B_s\to J/\psi \pi^+\pi^-$.  We  have derived  the $B_s\to \pi^+\pi^-$ and $D_s\to \pi^+\pi^-$ transition form factor using the light-cone sum rules, and then  presented  our results  for differential decay width which agree well with experimental data.   Accurate measurements by the  BES-III at the BEPC, the LHCb at the LHC and  Super-B factory at KEKB in future will be valuable   to more precisely  examine  our formalism, and determine the two-hadron LCDA.

\section*{Acknowledgements}
The authors are   grateful to  Zhi-Hui Guo, Hsiang-Nan Li, Cai-Dian L\"u,   Ulf-G. Mei{\ss}ner,  Wen-Fei Wang and Rui-Lin Zhu for  enlightening discussions. This work was supported in part  by Shanghai Natural  Science Foundation  under Grant  No. 11DZ2260700 and No. 15ZR1423100,  by the Open Project Program of State Key Laboratory of Theoretical Physics, Institute of Theoretical Physics, Chinese  Academy of Sciences, China (No.Y5KF111CJ1), and  by the Scientific Research Foundation for the Returned Overseas Chinese Scholars, State
Education Ministry.




\begin{thebibliography}{11}

\bibitem{Aaij:2013qta}
  R.~Aaij {\it et al.} [LHCb Collaboration],
  Phys.\ Rev.\ Lett.\  {\bf 111}, 191801 (2013)
  [arXiv:1308.1707 [hep-ex]].


\bibitem{LHCb:2015dla}
  The LHCb Collaboration [LHCb Collaboration],
  LHCb-CONF-2015-002, CERN-LHCb-CONF-2015-002.



\bibitem{Lu:2011jm}
  C.~D.~Lu and W.~Wang,
  Phys.\ Rev.\ D {\bf 85}, 034014 (2012)
  [arXiv:1111.1513 [hep-ph]].


\bibitem{Dey:2015rqa}
  B.~Dey,
  arXiv:1505.02873 [hep-ex].


\bibitem{Gratrex:2015hna}
  J.~Gratrex, M.~Hopfer and R.~Zwicky,
  arXiv:1506.03970 [hep-ph].


\bibitem{Becirevic:2012dp}
  D.~Becirevic and A.~Tayduganov,
  Nucl.\ Phys.\ B {\bf 868}, 368 (2013)
  [arXiv:1207.4004 [hep-ph]].


\bibitem{Matias:2012qz}
  J.~Matias,
  Phys.\ Rev.\ D {\bf 86}, 094024 (2012)
  [arXiv:1209.1525 [hep-ph]].

\bibitem{Blake:2012mb}
  T.~Blake, U.~Egede and A.~Shires,
  JHEP {\bf 1303}, 027 (2013)
  [arXiv:1210.5279 [hep-ph]].

\bibitem{Das:2014sra}
  D.~Das, G.~Hiller, M.~Jung and A.~Shires,
  JHEP {\bf 1409}, 109 (2014)
  [arXiv:1406.6681 [hep-ph]].

\bibitem{Hofer:2015kka}
  L.~Hofer and J.~Matias,
  arXiv:1502.00920 [hep-ph].

\bibitem{Das:2015pna}
  D.~Das, G.~Hiller and M.~Jung,
  arXiv:1506.06699 [hep-ph].

\bibitem{Meissner:2013hya}
  U.~G.~Meißner and W.~Wang,
  Phys.\ Lett.\ B {\bf 730}, 336 (2014)
  [arXiv:1312.3087 [hep-ph]].


\bibitem{Meissner:2013pba}
  U.~G.~Meißner and W.~Wang,
  JHEP {\bf 1401}, 107 (2014)
  [arXiv:1311.5420 [hep-ph]].


\bibitem{Doring:2013wka}
  M.~Döring, U.~G.~Meißner and W.~Wang,
  JHEP {\bf 1310}, 011 (2013)
  [arXiv:1307.0947 [hep-ph]].


\bibitem{Wang:2014sba}
  W.~Wang,
  Int.\ J.\ Mod.\ Phys.\ A {\bf 29}, 1430040 (2014)
  [arXiv:1407.6868 [hep-ph]].


\bibitem{Gardner:2001gc}
  S.~Gardner and U.~G.~Meissner,
  Phys.\ Rev.\ D {\bf 65}, 094004 (2002)
  [hep-ph/0112281].


\bibitem{Maul:2001zn}
  M.~Maul,
  Eur.\ Phys.\ J.\ C {\bf 21}, 115 (2001)
  [hep-ph/0104078].



\bibitem{Liang:2014tia}
  W.~H.~Liang and E.~Oset,
  Phys.\ Lett.\ B {\bf 737}, 70 (2014)
  [arXiv:1406.7228 [hep-ph]].


\bibitem{Bayar:2014qha}
  M.~Bayar, W.~H.~Liang and E.~Oset,
  Phys.\ Rev.\ D {\bf 90}, 114004 (2014)
  [arXiv:1408.6920 [hep-ph]].


\bibitem{Xie:2014gla}
  J.~J.~Xie and E.~Oset,
  Phys.\ Rev.\ D {\bf 90}, 094006 (2014)
  [arXiv:1409.1341 [hep-ph]].


\bibitem{Sayahi:2013tza}
  M.~Sayahi and H.~Mehraban,
  Phys.\ Scripta {\bf 88}, 035101 (2013).

\bibitem{Sekihara:2015iha}
  T.~Sekihara and E.~Oset,
  arXiv:1507.02026 [hep-ph].


\bibitem{Roca:2015tea}
  L.~Roca, M.~Mai, E.~Oset and U.~G.~Meißner,
  Eur.\ Phys.\ J.\ C {\bf 75}, no. 5, 218 (2015)
  [arXiv:1503.02936 [hep-ph]].


\bibitem{Chen:2002th}
  C.~H.~Chen and H.~n.~Li,
  Phys.\ Lett.\ B {\bf 561}, 258 (2003)
  [hep-ph/0209043].


\bibitem{Chen:2004az}
  C.~H.~Chen and H.~n.~Li,
  Phys.\ Rev.\ D {\bf 70}, 054006 (2004)
  [hep-ph/0404097].


\bibitem{Wang:2014ira}
  W.~F.~Wang, H.~C.~Hu, H.~n.~Li and C.~D.~L?
  Phys.\ Rev.\ D {\bf 89}, no. 7, 074031 (2014)
  [arXiv:1402.5280 [hep-ph]].


\bibitem{Wang:2014qya}
  H.~s.~Wang, S.~m.~Liu, J.~Cao, X.~Liu and Z.~j.~Xiao,
  Nucl.\ Phys.\ A {\bf 930}, 117 (2014).


\bibitem{Aaij:2014lba}
  R.~Aaij {\it et al.} [LHCb Collaboration],
  Phys.\ Lett.\ B {\bf 743}, 46 (2015)
  [arXiv:1412.6433 [hep-ex]].


\bibitem{Wang:2015uea}
  W.~F.~Wang, H.~n.~Li, W.~Wang and C.~D.~L?
  Phys.\ Rev.\ D {\bf 91}, no. 9, 094024 (2015)
  [arXiv:1502.05483 [hep-ph]].


\bibitem{Wang:2015paa}
  W.~Wang and R.~L.~Zhu,
  Phys.\ Lett.\ B {\bf 743}, 467 (2015)
  [arXiv:1502.05104 [hep-ph]].


\bibitem{Yelton:2009aa}
  J.~Yelton {\it et al.} [CLEO Collaboration],
  Phys.\ Rev.\ D {\bf 80}, 052007 (2009)
  [arXiv:0903.0601 [hep-ex]].


\bibitem{Ecklund:2009aa}
  K.~M.~Ecklund {\it et al.} [CLEO Collaboration],
  Phys.\ Rev.\ D {\bf 80}, 052009 (2009)
  [arXiv:0907.3201 [hep-ex]].


\bibitem{Hietala:2015jqa}
  J.~Hietala, D.~Cronin-Hennessy, T.~Pedlar and I.~Shipsey,
  Phys.\ Rev.\ D {\bf 92}, no. 1, 012009 (2015)
  [arXiv:1505.04205 [hep-ex]].


\bibitem{Asner:2008nq}
  D.~M.~Asner {\it et al.},
  Int.\ J.\ Mod.\ Phys.\ A {\bf 24}, S1 (2009)
  [arXiv:0809.1869 [hep-ex]].


\bibitem{Colangelo:2000dp}
  P.~Colangelo and A.~Khodjamirian,
  In *Shifman, M. (ed.): At the frontier of particle physics, vol. 3* 1495-1576
  [hep-ph/0010175].


\bibitem{Gasser:1990bv}
  J.~Gasser and U.~G.~Meissner,
  Nucl.\ Phys.\ B {\bf 357}, 90 (1991).


\bibitem{Meissner:2000bc}
  U.~G.~Meissner and J.~A.~Oller,
  Nucl.\ Phys.\ A {\bf 679}, 671 (2001)
  [hep-ph/0005253].


\bibitem{Bijnens:2003uy}
  J.~Bijnens and P.~Talavera,
  Nucl.\ Phys.\ B {\bf 669}, 341 (2003)
  [hep-ph/0303103].


\bibitem{Lahde:2006wr}
  T.~A.~Lahde and U.~G.~Meissner,
  Phys.\ Rev.\ D {\bf 74}, 034021 (2006)
  [hep-ph/0606133].


\bibitem{Guo:2012yt}
  Z.~H.~Guo, J.~A.~Oller and J.~Ruiz de Elvira,
  Phys.\ Rev.\ D {\bf 86}, 054006 (2012)
  [arXiv:1206.4163 [hep-ph]].


\bibitem{Gasser:1983yg}
  J.~Gasser and H.~Leutwyler,
  Annals Phys.\  {\bf 158}, 142 (1984).


\bibitem{Gasser:1984gg}
  J.~Gasser and H.~Leutwyler,
  Nucl.\ Phys.\ B {\bf 250}, 465 (1985).


\bibitem{Gasser:1984ux}
  J.~Gasser and H.~Leutwyler,
  Nucl.\ Phys.\ B {\bf 250}, 517 (1985).


\bibitem{Donoghue:1990xh}
  J.~F.~Donoghue, J.~Gasser and H.~Leutwyler,
  Nucl.\ Phys.\ B {\bf 343}, 341 (1990).


\bibitem{Oller:1998hw}
  J.~A.~Oller, E.~Oset and J.~R.~Pelaez,
  Phys.\ Rev.\ D {\bf 59}, 074001 (1999)
  [Phys.\ Rev.\ D {\bf 60}, 099906 (1999)]
  [Phys.\ Rev.\ D {\bf 75}, 099903 (2007)]
  [hep-ph/9804209].


\bibitem{Ablikim:2004wn}
  M.~Ablikim {\it et al.} [BES Collaboration],
  Phys.\ Lett.\ B {\bf 607}, 243 (2005)
  [hep-ex/0411001].


\bibitem{Aaij:2013zpt}
  R.~Aaij {\it et al.} [LHCb Collaboration],
  Phys.\ Rev.\ D {\bf 87}, no. 5, 052001 (2013)
  [arXiv:1301.5347 [hep-ex]].


\bibitem{Aaij:2014emv}
  R.~Aaij {\it et al.} [LHCb Collaboration],
  Phys.\ Rev.\ D {\bf 89}, no. 9, 092006 (2014)
  [arXiv:1402.6248 [hep-ex]].


\bibitem{Flatte:1976xu}
  S.~M.~Flatte,
  Phys.\ Lett.\ B {\bf 63}, 224 (1976).


\bibitem{Flatte:1976xv}
  S.~M.~Flatte,
  Phys.\ Lett.\ B {\bf 63}, 228 (1976).


\bibitem{Diehl:1998dk}
  M.~Diehl, T.~Gousset, B.~Pire and O.~Teryaev,
  Phys.\ Rev.\ Lett.\  {\bf 81}, 1782 (1998)
  [hep-ph/9805380].


\bibitem{Polyakov:1998ze}
  M.~V.~Polyakov,
  Nucl.\ Phys.\ B {\bf 555}, 231 (1999)
  [hep-ph/9809483].


\bibitem{Kivel:1999sd}
  N.~Kivel, L.~Mankiewicz and M.~V.~Polyakov,
  Phys.\ Lett.\ B {\bf 467}, 263 (1999)
  [hep-ph/9908334].


\bibitem{Diehl:2003ny}
  M.~Diehl,
  Phys.\ Rept.\  {\bf 388}, 41 (2003)
  [hep-ph/0307382].


\bibitem{Braun:2003rp}
  V.~M.~Braun, G.~P.~Korchemsky and D.~Mueller,
  Prog.\ Part.\ Nucl.\ Phys.\  {\bf 51}, 311 (2003)
  [hep-ph/0306057].


\bibitem{Cheng:2005nb}
  H.~Y.~Cheng, C.~K.~Chua and K.~C.~Yang,
  Phys.\ Rev.\ D {\bf 73}, 014017 (2006)
  [hep-ph/0508104].


\bibitem{Mueller:1998fv}
  D.~Müller, D.~Robaschik, B.~Geyer, F.-M.~Dittes and J.~Ho?ej¨i,
  Fortsch.\ Phys.\  {\bf 42}, 101 (1994)
  [hep-ph/9812448].


\bibitem{Hagler:2002nh}
  P.~Hagler, B.~Pire, L.~Szymanowski and O.~V.~Teryaev,
  Phys.\ Lett.\ B {\bf 535}, 117 (2002)
  [Phys.\ Lett.\ B {\bf 540}, 324 (2002)]
  [hep-ph/0202231].


\bibitem{Pire:2008xe}
  B.~Pire, F.~Schwennsen, L.~Szymanowski and S.~Wallon,
  Phys.\ Rev.\ D {\bf 78}, 094009 (2008)
  [arXiv:0810.3817 [hep-ph]].


\bibitem{Watson:1952ji}
  K.~M.~Watson,
  Phys.\ Rev.\  {\bf 88}, 1163 (1952).


\bibitem{Migdal:1955}
  A.~B.~Migdal,
 Sov. Phys. JETP,   {\bf 1}, 2 (1955).

\bibitem{Aoki:2013ldr}
  S.~Aoki {\it et al.},
  Eur.\ Phys.\ J.\ C {\bf 74}, 2890 (2014)
  [arXiv:1310.8555 [hep-lat]].


\bibitem{Agashe:2014kda}
  K.~A.~Olive {\it et al.} [Particle Data Group Collaboration],
  Chin.\ Phys.\ C {\bf 38}, 090001 (2014).


\bibitem{Colangelo:2010bg}
  P.~Colangelo, F.~De Fazio and W.~Wang,
  Phys.\ Rev.\ D {\bf 81}, 074001 (2010)
  [arXiv:1002.2880 [hep-ph]].


\bibitem{Buchalla:1995vs}
  G.~Buchalla, A.~J.~Buras and M.~E.~Lautenbacher,
  Rev.\ Mod.\ Phys.\  {\bf 68}, 1125 (1996)
  [hep-ph/9512380].



\bibitem{Bediaga:2012py}
  R.~Aaij {\it et al.} [LHCb Collaboration],
  Eur.\ Phys.\ J.\ C {\bf 73}, no. 4, 2373 (2013)
  [arXiv:1208.3355 [hep-ex]].


\bibitem{Aushev:2010bq}
  T.~Aushev {\it et al.},
  arXiv:1002.5012 [hep-ex].


\bibitem{Inami:1980fz}
  T.~Inami and C.~S.~Lim,
  Prog.\ Theor.\ Phys.\  {\bf 65}, 297 (1981)
  [Prog.\ Theor.\ Phys.\  {\bf 65}, 1772 (1981)].


\bibitem{Buchalla:1993bv}
  G.~Buchalla and A.~J.~Buras,
  Nucl.\ Phys.\ B {\bf 400}, 225 (1993).


\bibitem{Buchalla:1998ba}
  G.~Buchalla and A.~J.~Buras,
  Nucl.\ Phys.\ B {\bf 548}, 309 (1999)
  [hep-ph/9901288].


\bibitem{Misiak:1999yg}
  M.~Misiak and J.~Urban,
  Phys.\ Lett.\ B {\bf 451}, 161 (1999)
  [hep-ph/9901278].




\end{thebibliography}
\end{document}